\documentclass{iaus}
\usepackage{graphicx}

\title[Tidal-pulsational interaction in FN\,CMa] 
      {The (B0$+$?)$+$O6 system FN\,CMa\footnote{Based on data from
          ESO programs 076.C-0431, 076.C-0164, and the BESO spectrograph.}:
        \\ A case for tidal-pulsational interaction?}

\author[Th.\ Rivinius, O.\ Stahl, S.\ {\v S}tefl et al.]   
{Th.\ Rivinius$^{1}$, 
O.\ Stahl$^2$,
S.\ {\v S}tefl$^1$,
D.\ Baade$^3$,
R.H.D.\ Townsend$^4$,
\\
L.\ Barrera$^5$
}

\affiliation{$^1$ESO Chile; 
$^2$LSW/ZAH Heidelberg, Germany; 
$^3$ESO Germany; 
$^4$UW Madison, USA;
$^5$UMCE Santiago, Chile
}

\pubyear{2011}
\volume{272}  
\pagerange{1--2}
\jname{Active OB stars: structure, evolution, mass loss and critical limits}
\editors{C.\ Neiner, G.\ Wade, G.\ Meynet \& G.\ Peters, eds.}
\begin{document}
\maketitle

\begin{abstract}
FN\,CMa is visually double with a separation of $\sim$0.6\,arcsec.  Sixty
high-cadence VLT/{\it UVES} spectra permit the A and B components to be
disentangled, as the relative contribution of each star to the total light
entering the spectrograph fluctuates between exposures due to changes in
seeing. Component A exhibits rapid line-profile variations, leading us to
attribute the photometric variability seen by HIPPARCOS (with a derived
$P=0.08866$\,d) to this component.  From a total of 122 archival and new
echelle spectra it is shown that component A is an SB1 binary with an orbital
period of 117.55 days.  The eccentricity of 0.6 may result in tidal modulation
of the pulsation(s) of component Aa.  
\end{abstract}

\firstsection 
\section{Introducing FN\,CMa}
FN\,CMa (HD\,53\,974) is a bright ($V=5.4$\,mag) B0.5\,III star and
visually double.  Within about a century, the relative position of
components A and B, which are separated by $\sim$0.6 arcsec, has
changed marginally at most.  A is brighter than B by about 1.2 mag.

\section{Observations and data reduction}
The ESO Science Archive contains 60 VLT/{\it UVES} echelle spectra of FN\,CMa
obtained within 1.4 hours for a study of interstellar medium, and three more
spectra from {\it FEROS} at the 2.2-m ESO/MPG telescope, La Silla.  In 2009
and 2010, an additional 59 echelle spectra were secured with the {\it BESO}
spectrograph, a clone of {\it FEROS} mounted on the Bochum 1.5-m Hexapod
Telescope on Cerro Armazones.

As a result of variable seeing and imperfect guiding, some UVES spectra
contain a significantly higher fraction of light from component B than
others. Since the light combination is geometric it has no spectral
dependency, and thus, under the assumption that certain spectral features are
due to either A (e.g., Si\,{\sc iii} 4553) or B (e.g., He\,{\sc ii} 4540)
alone, a simple linear set of equations can be used for the disentangling of
the spectra from the two stars over the entire wavelength range.  The
inferred spectral light ratios, between 0.75 and 0.85, are in good agreement
with the known magnitude difference.

\section{Results}
{\bf FN\,CMa B:} This component has a spectrum typical of mid-O main-sequence
stars.  Compared to the B0.5III primary, it would be considerably
underluminous if the pair were physical.  However, assuming an O subdwarf
companion does not help because, then, component B would be about 2 mag {\it
  over}\/luminous.
Components A and B display the same set of interstellar Ca\,{sc ii} K lines
except that the redmost one is significantly stronger in B.  Considering also
incipient emission in N\,{\sc iii} and H$\alpha$, we conclude that component B
is best described as an O6V((f)) background star.

\begin{figure}
\centering
\parbox{\textwidth}{%
\parbox{.52\textwidth}{\includegraphics[angle=270,width=.52\textwidth,clip]{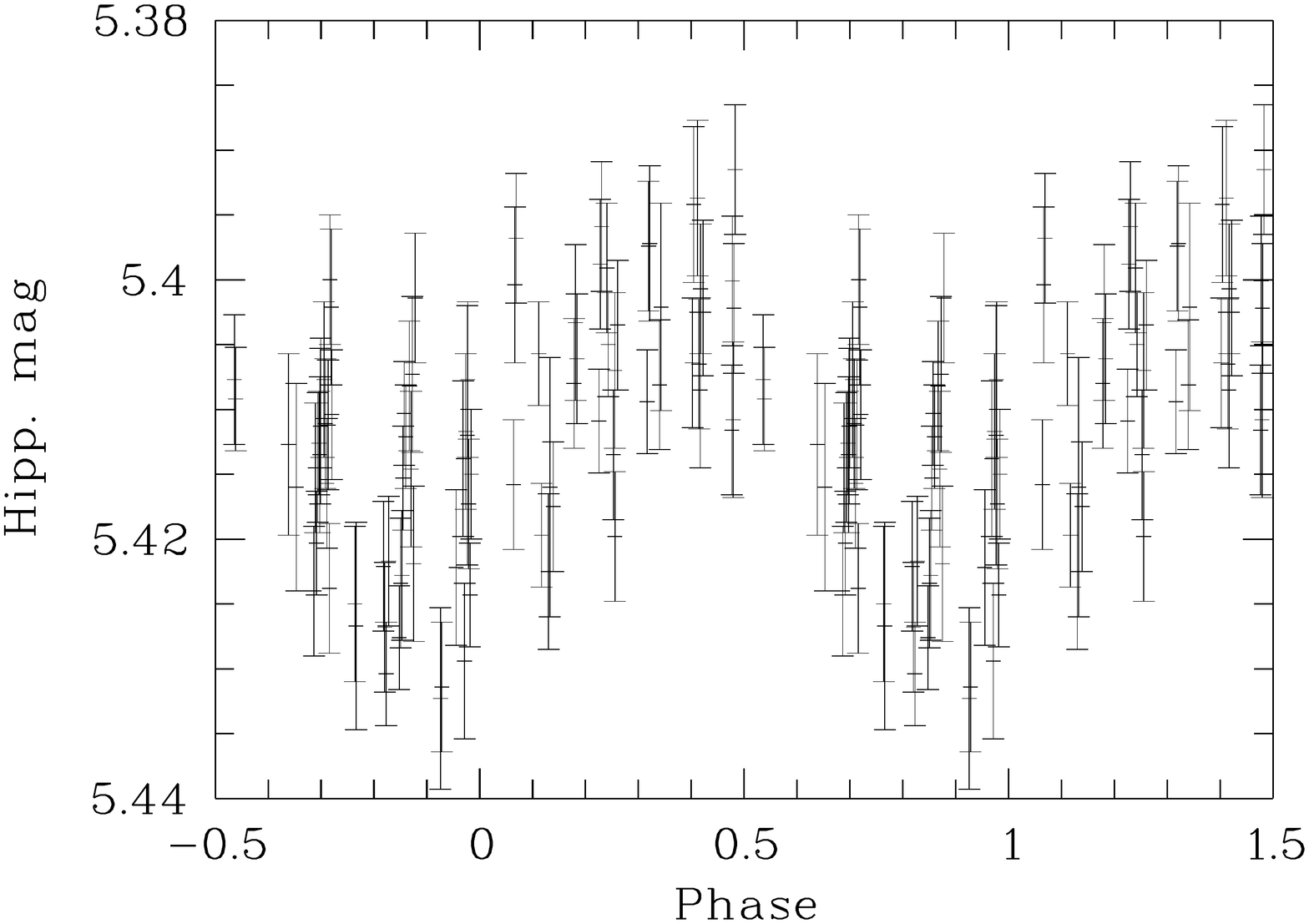}}%
\parbox{.48\textwidth}{\includegraphics[viewport=15 18 694 500,angle=0,width=.48\textwidth,clip]{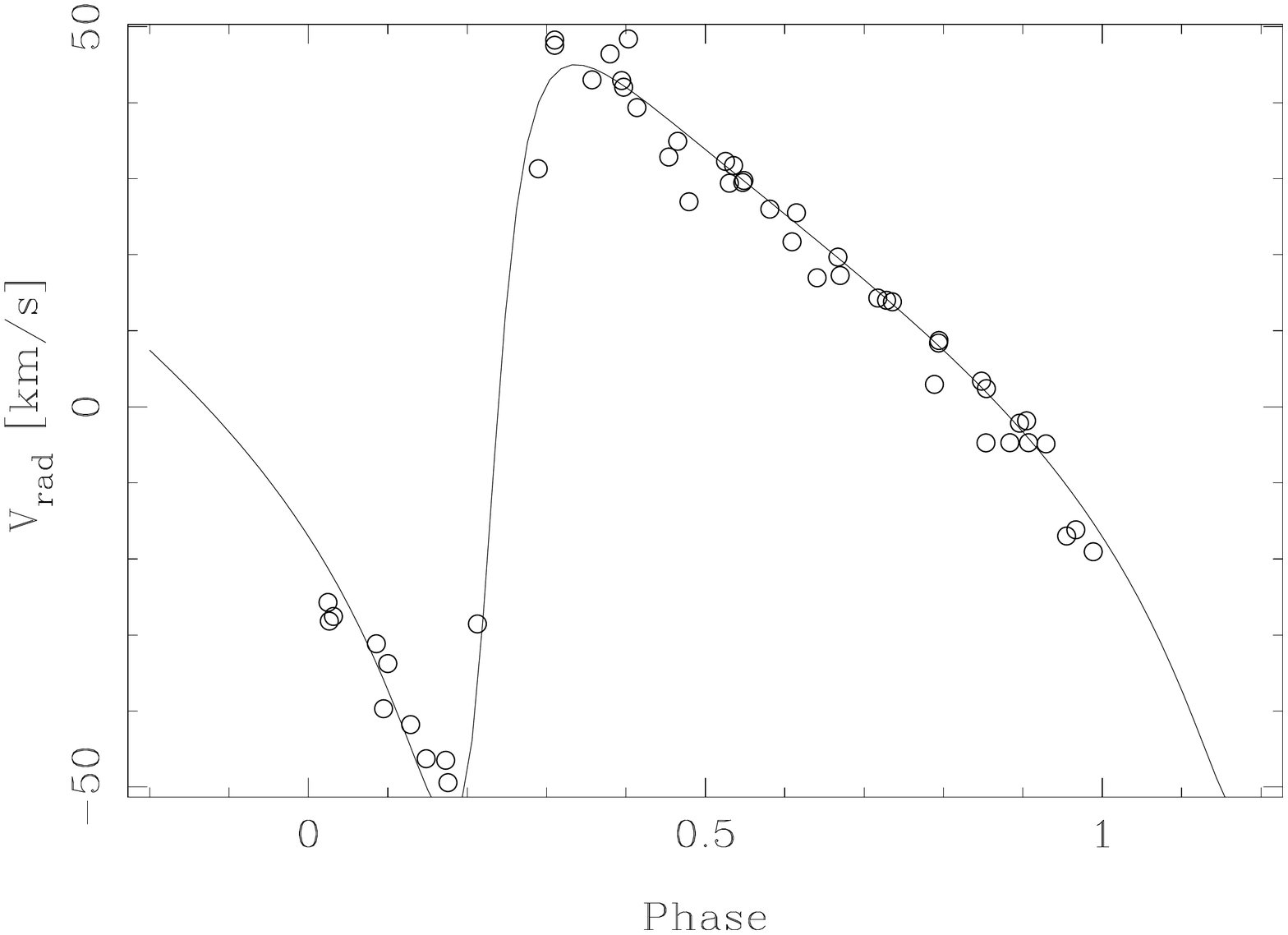}}%
}
\caption[]{
\centering
Left panel: {\it HIPPARCOS} photometry of FN\,CMa phased with a period
of 2.13\,h. \newline Right panel: The radial velocity curve of FN\,CMa\,Aa
(P\,=\,117.55\,d).}
\label{FNCMa}
\end{figure}

{\bf FN\,CMa A:} In the literature, FN\,CMa has a record of low-amplitude
photometric variability and modulated spectral line profiles.  But there is no
consensus about its nature.  Our analysis of the {\it HIPPARCOS} photometry
yields a period of 0.08866\,d (2.13\,h; see left panel of Fig.\ \ref{FNCMa})
with $\sim$0.02\,mag amplitude.  The combination of spectral type, period, and
amplitude makes FN\,CMa a $\beta$ Cephei star candidate, as already suggested
by other observers.  This is further supported by the rapid spectral
line-profile variability of component A (however, at just 1.4\,h, the {\it
  UVES} data string is too short and the {\it BESO} spectra are not
sufficiently densely sampled to attempt an independent period determination).
In any case, given the spectral variations seen in component A, we attribute
the photometric variability to this component as well.

Much larger-amplitude long-term radial-velocity variability is apparent from
the {\it BESO} data: FN\,CMa\,A is itself an SB1 binary with the following
properties:
\vspace*{5mm}

\noindent{\footnotesize
\begin{center}
\begin{tabular}{lr}
Period [d]			&  117.55 $\pm$ 0.33		\\
Periastron epoch [JD]		&  2\,453\,779.5 $\pm$ 4	\\
Periastron longitude [deg]	&  247 $\pm$ 7			\\
$e$				&  0.60 $\pm$ 0.05		\\
$K_1$ [km/s]			&  49.8 $\pm$ 3.5		\\
$\gamma$ [km/s]			&  5.9  $\pm$ 1.5		\\
\end{tabular}
\end{center}}
\vspace*{5mm}

\noindent
The radial-velocity curve of FN\,CMa\,Aa is shown in Fig.\ \ref{FNCMa} (right
panel).  Its relatively large amplitude suggests that the so-far (directly)
undetected component FN\,CM\,Ab is a fairly massive star. However, it appears
too faint to be the carrier of the rapid variability.

\section{Discussion}

The high eccentricity and moderate orbital period of the subsystem
FN\,CMa\,Aa$+$Ab may enable searches for a tidal modulation of the
pulsation of component Aa.  Since FN\,CMa is bright and situated in a
region with numerous other pulsating OB stars, it might be worthwhile
including it in the target lists of wide-angle asteroseismology
satellites such as BRITE.

\end{document}